\documentclass[a4 paper, 12 pt] {article}

\usepackage{graphics, graphicx}

\begin{document}
\title{\bf{Integrability and Hopf Solitons in Models with Explicitly Broken
$O(3)$ Symmetry }}
\author{A. Wereszczy\'{n}ski $^{a)}$
\thanks{wereszczynski@th.if.uj.edu.pl}
       \\
       \\ $^{a)}$ Institute of Physics,  Jagiellonian University,
       \\ Reymonta 4, Krak\'{o}w, Poland}
\maketitle
\begin{abstract}
A wide class of models, built of the three component unit vector
field living in the (3+1) Minkowski space-time, which break
explicitly global $O(3)$ symmetry are discussed. The symmetry
breaking occurs due to the so-called dielectric function
multiplying a standard symmetric term. Integrability conditions
are found. Moreover, for some particular forms of the Lagrangian
exact toroidal solutions with any Hopf index are obtained. It is
proved that such symmetry breaking influences the shape of the
solitons whereas the energy as well as the Hopf index remain
unchanged.
\end{abstract}
\newpage
\section{\bf{ Introduction}}
Since Faddeev, Niemi \cite{niemi}, \cite{langmann} and Cho
\cite{cho} have proposed their famous action, which is considered
as a good candidate for an effective model describing the
non-perturbative regime of the gluodynamics, toroidal topological
field configurations i.e. toroidal solitons have been widely
investigated \cite{nicole}, \cite{battyde}, \cite{salo},
\cite{ward}, \cite{vanbaal}. They give a very attractive framework
to understand the physics of the glueballs - particles maid only
of the gauge field. In this picture a particular glueball can be
described by a toroidal soliton with the pertinent topological
number, so-called Hopf index \cite{niemi_new}, \cite{hopfions}. It
is in the agreement with the standard picture of the mesons, where
quarks are connected by a very thin tube of the gauge field. Now,
because of the fact that glueballs do not consist of quarks, such
a flux-tube cannot end on the sources. In order to form a stable
object the ends must be joined giving loop-like configurations.
\\
However, in spite of the fact that Faddeev-Niemi model gives us
the chance for the very elegant way to describe physics of the
glueballs, it has also its own problems. One of the most important
problems is the existence of massless excitations, that is
Goldstone bosons which appear as the effect of the spontaneous
symmetry breaking. In fact, the Lagrangian has global $O(3)$
symmetry whereas the vacuum state is only $O(2)$ invariant. Thus,
two generators are broken and two bosons emerge. This feature of
Faddeev-Niemi model has been recently discussed and some
modifications have been proposed \cite{niemi2}, \cite{sanchez1},
\cite{wipf}, \cite{my}. In general, they mean necessity of the
adding new terms into the Lagrangian, which assure the explicit
$O(3)$ symmetry breaking.
\\
The main aim of the present work is to analyze toroidal solitons
in models with explicitly broken $O(3)$ symmetry. Such problems as
existence of knotted solitons in these models, influence of
symmetry breaking term on energy and shape of knots have been not
analyzed yet. In this paper it will be shown that unbroken $O(3)$
global symmetry is not necessary condition for existence of knots.
Of course, there are infinitely many way in which the symmetry can
be broken at the Lagrangian level. Here, the simplest case is
chosen - a symmetric part of the model is multiplied by some
function, usually called dielectric function, which explicitly
breaks the symmetry. However, such simple pattern of the symmetry
breaking is strongly motivated by the latest investigation of the
effective action for the low energy QCD. Namely, using the
non-Abelian color dielectric model one can derive, in the limit
when the color dielectric field condenses, a modified version of
the Faddeev-Niemi action \cite{my}, \cite{arodz}. The modification
is given by two functions which multiply the standard
Faddeev-Skyrme and kinetic terms. These functions depend only on
the unit vector field and break explicitly $O(3)$ symmetry. One
can immediately ask about the fate of the toroidal solutions in
the theories with explicitly broken $O(3)$ symmetry.
Unfortunately, obtained model is even more complicated than the
standard Faddeev-Niemi model and none analytical solutions are
known. In spite of that, this problem can be investigated and
solved in case of the toy model which is defined below.
\\
To conclude, the question how the explicit breaking of the global
$O(3)$ symmetry modifies the topological toroidal solitons
deserves answer. It is clear that results obtained here do not
have to concern the modified versions of the Faddeev-Niemi model.
Nonetheless, some effects seem to be quite general and might
appear in some modified models \cite{my} as well. Thus, one can
treat this paper as the first step to understand what happen with
hopfions in the modified Faddeev-Niemi theories.
\section{\bf{ The Model}}
Let us start with the Aratyn-Ferreira-Zimerman model
\cite{aratyn}, \cite{sanchez2}
\begin{equation}
\mathcal{L}= \left[ [ \vec{n} \cdot (
\partial_{\mu } \vec{n} \times
\partial_{\nu } \vec{n} )]^2 \right]^{\frac{3}{4}},
\label{arratyn}
\end{equation}
where $\vec{n}$ is a three component unit vector leaving in the
Minkowski space-time. \\
In our work we generalize it to the following Lagrangian density
\begin{equation}
\mathcal{L}= \sigma \left(\vec{n} \right) \left[ [ \vec{n} \cdot (
\partial_{\mu } \vec{n} \times
\partial_{\nu } \vec{n} )]^2 \right]^{\frac{3}{4}}.
\label{toylag}
\end{equation}
The symmetry breaking occurs due to the so-called dielectric
function $\sigma $, which depends on the unit field only. On the
other hand, the symmetric part can be written as a function of
antisymmetric field tensor $H_{\mu \nu} =\vec{n} \cdot (
\partial_{\mu } \vec{n} \times \partial_{\nu } \vec{n} )$. The
unit field can be, in the standard way, expressed by a complex
field $u, \, u^*$:
\begin{equation}
\vec{n}= \frac{1}{1+|u|^2} ( u+u^*, -i(u-u^*), |u|^2-1).
\label{stereograf}
\end{equation}
Then we obtain
\begin{equation}
\mathcal{L}= \sigma \left( u,u^* \right)
\frac{8^{3/4}}{(1+|u|^2)^3} (K_{\mu} \partial^{\mu}
u^*)^{\frac{3}{4}}, \label{toylag1}
\end{equation}
where following \cite{aratyn} we define $K_{\mu} =(\partial_{\nu}
u^*
\partial^{\nu} u)
\partial_{\mu} u -(\partial_{\nu}u \partial^{\nu } u)
\partial_{\mu}u^*$. This object has two important properties:
\begin{equation}
K_{\mu} \partial^{\mu} u=0 \; \; \; \; \; \mbox{and} \; \; \; \;
\; Im(K_{\mu}
\partial^{\mu} u^* )=0. \label{properties}
\end{equation}
The pertinent equations of motion read
\begin{equation}
\partial_{\mu} \frac{3}{2}\left[  \frac{\sigma}{(1+|u|^2)^3}
(K_{\mu} \partial^{\mu} u^*)^{-\frac{1}{4}} K^{\mu} \right] -
(K_{\mu} \partial^{\mu} u^*)^{\frac{3}{4}} \left[
\frac{\sigma}{(1+|u|^2)^3} \right]'_{u^*} =0 \label{eqmot-toy1}
\end{equation}
and its complex conjugation. One can rewrite it into the more
compact form
\begin{equation}
\partial_{\mu} \left[  \frac{\sigma^{1/3}}{(1+|u|^2)}
(K_{\mu} \partial^{\mu} u^*)^{-\frac{1}{4}} K^{\mu} \right]=0,
\label{eqmot-toy2}
\end{equation}
or
\begin{equation}
\partial_{\mu} \mathcal{K}^{\mu} =0, \label{eqmot-toy3}
\end{equation}
where
\begin{equation}
\mathcal{K}_{\mu} = \frac{\sigma^{1/3}}{(1+|u|^2)} (K_{\mu}
\partial^{\mu} u^*)^{-\frac{1}{4}} K^{\mu}. \label{int_K}
\end{equation}
One can check that $\mathcal{K}$ fulfills the properties
(\ref{properties}) only if the dielectric function is a real
function $$\sigma = \sigma^*.$$ After that we are able to define
an infinite family of the conserved currents using the procedure
proposed in \cite{aratyn}. Namely,
\begin{equation}
J_{\mu} \equiv \mathcal{K}_{\mu} \frac{ \partial G}{\partial u}
-\mathcal{K}_{\mu}^* \frac{\partial G}{\partial u^*},
\label{int_current}
\end{equation}
where $G=G(u,u^*)$. Thus, the analyzed model is integrable.
Integrability is understood as existence of the infinite number of
the conserved currents \cite{alvarez}, \cite{ferreira}. It is
straightforward to check that the integrability is observed for
all models where the symmetric part is any function of the field
tensor $H_{\mu \nu}^2$. The case considered here has been chosen
to omit the Derick theorem for non-existence of stable solitons.
Thus, we see that integrability property is not connected with the
global $O(3)$ symmetry. It is rather unexpected result. Usually,
integrability (and solitons) are observed in maximally symmetric
situation.
\\
Now, we show that, as in the standard theories with solitons,
integrability leads to appearance of soliton solutions.
\\
Because of the fact that our aim is to find topological toroidal
solitons with the non-trivial Hopf number we introduce the
toroidal coordinates
$$ x=\frac{a}{q} \sinh \eta \cos \phi , $$
$$ y=\frac{a}{q} \sinh \eta \sin \phi , $$
\begin{equation}
z=\frac{a}{q} \sin \xi ,\label{tor_coord}
\end{equation}
where $q=\cosh \eta -\cos \xi $ and $a>0$ is a constant of
dimension of length fixing the scale in the coordinates. Moreover,
we assume the following Ansatz for the field $u$ \cite{aratyn}
\begin{equation}
u(\eta,\xi,\phi) \equiv f(\eta) e^{i(m\xi + n \phi)},
\label{anzatz}
\end{equation}
where $m, \; n$ are integers.
\\
Then, the static equation of motion takes the form
\begin{equation}
\partial_{\eta} \ln \frac{\sigma^{2/3} ff'}{(1+f^2)^2} =-\frac{2m^2 \sinh^2 \eta -n^2}{m^2 \sinh^2 \eta
+n^2} \frac{\cosh \eta}{\sinh \eta}. \label{eqmot-toy4a}
\end{equation}
It can be integrated and we find that
\begin{equation}
\frac{\sigma^{2/3} ff'}{(1+f^2)^2} = \frac{k_1}{|m|^3} \frac{\sinh
\eta}{\left( \frac{n^2-m^2}{m^2} +\cosh^2 \eta \right)^{3/2}},
\label{eqmot-toy4}
\end{equation}
where $k_1$ is a constant.
\\
Finally for any dielectric function, we obtain the general
solution given by the integral
\begin{equation}
\int \frac{\sigma^{2/3} f}{(1+f^2)^2} df=
\frac{-k_1}{|m|(m^2-n^2)} \frac{\cosh \eta}{\left(
\frac{n^2-m^2}{m^2} +\cosh^2 \eta \right)^{1/2}} -\frac{k_2}{2},
\label{eqmot-toy5}
\end{equation}
where $k_2$ is the second integration constant. Here, the case
$|m|>|n|$ has been assumed.
\\
Before we specify the dielectric function and present exact
solution let us consider the total energy of the solution given by
(\ref{eqmot-toy5}). In case of our model, the total energy reads
\begin{equation}
E \equiv \int d^3x T_{00} = 8^{3/4} \int d^3x \sigma(u,u^*)
\frac{(K_i \partial^i u^*)^{\frac{3}{4}}}{ (1+|u|^2)^3}.
\label{energy1}
\end{equation}
Using (\ref{anzatz}) we get
\begin{equation}
E_{m,n}=(2\pi)^2 8 \cdot 2^{3/4} \int_0^{\infty} \frac{d \eta
\sinh \eta}{(1+f^2)^{3}} \left( m^2 +\frac{n^2}{\sinh^2 \eta }
\right)^{\frac{3}{4}} f^{\frac{3}{2}} f'^{\frac{3}{2}} \sigma (f).
\label{energy2}
\end{equation}
Quite surprisingly this expression can be integrated for any
dielectric function. In fact, using equation (\ref{eqmot-toy4}) we
are able to remove $\sigma $ from (\ref{energy2}). Then
\begin{equation}
E_{m,n}=(2\pi)^2 8 \cdot 2^{3/4} |k_1|^{\frac{3}{2}}
\int_0^{\infty} \frac{d \eta }{\sinh^{2} \eta } \left( m^2
+\frac{n^2}{\sinh^2 \eta } \right)^{-\frac{3}{2}}. \label{energy3}
\end{equation}
Finally, we obtain that
\begin{equation}
E_{m,n}=(2\pi)^2 8 \cdot 2^{3/4} \frac{|k_1|^{\frac{3}{2}}
}{|m||n|(|m|+|n|)}.
\end{equation}
The total energy is given only in terms of the integer numbers
$m,n$ and the first integration constant $k_1$. This expression
strongly depends on the asymptotic value of the function $f$ i.e.
behavior of the unit vector field in origin and at the spatial
infinity.
\\
In order to find exact solution one has to choice a particular
form of the dielectric function and fix the asymptotic conditions.
Let us consider the following family of functions label by the
continuous parameter $\delta
>-\frac{3}{2}$
\begin{equation}
\sigma(\vec{n}) = (1-n^3)^{\delta} \alpha (\delta).
\label{example1}
\end{equation}
The new constant is chosen in the form
$$\alpha (\delta )=\frac{1}{2^{\delta}} \left( \frac{3+2\delta}{3}
\right)^{\frac{3}{2}}.$$ One can notice that a very similar
dielectric function appears in the non-Abelian color dielectric
model in the limit where dielectric field condenses on some
non-trivial value \cite{my}.
\\
This dielectric function can be expressed in terms of $f$ as
\begin{equation}
\sigma = \alpha (\delta )\left( \frac{2}{(1+f^2)}
\right)^{\delta}. \label{example2}
\end{equation}
Then, the integral (\ref{eqmot-toy5}) can be evaluated and we find
\begin{equation}
\frac{1}{1+f^2}=\left[ \left( \frac{2k_1}{|m|(m^2-n^2)}
\frac{\cosh \eta}{\left( \frac{n^2-m^2}{m^2} +\cosh^2 \eta
\right)^{1/2}} + k_2 \right) \right]^{\frac{3}{2\delta +3}}.
\label{sol1}
\end{equation}
To calculate unknown constants $k_1, k_2$ one has to fix the
asymptotic conditions. We take
\begin{equation}
\vec{n} \rightarrow (0,0,1) \; \; \mbox{i.e.} \; \; f \rightarrow
\infty \; \; \mbox{as} \; \; \eta \rightarrow 0 \label{bound1}
\end{equation}
and
\begin{equation}
\vec{n} \rightarrow (0,0,-1) \; \; \mbox{i.e.} \; \; f \rightarrow
0 \; \; \mbox{as} \; \; \eta \rightarrow \infty. \label{bound2}
\end{equation}
Thus, after some simply algebra one can obtain
\begin{equation}
k_1=\frac{1}{2} (m^2-n^2) \frac{|m||n|}{|n|-|m|} \label{k1}
\end{equation}
and
\begin{equation}
k_2= \frac{|m|}{|m|-|n|}. \label{k2}
\end{equation}
Inserting the constants in (\ref{sol1}) we derive that
\begin{equation}
\frac{1}{1+f^2}=\left[  \frac{1}{|m|-|n|} \left( |m| - |n|
\frac{\cosh \eta}{\left( \frac{n^2}{m^2} +\sinh^2 \eta
\right)^{1/2}} \right)
 \right]^{\frac{3}{2\delta +3}}. \label{sol2}
\end{equation}
Finally, the function $f$ is found to be
\begin{equation}
f^2=\frac{ \left( \left(\frac{|m|}{|n|} -1\right)
\sqrt{\frac{n^2}{m^2} +\sinh^2 \eta} \right)^{\frac{3}{2\delta
+3}} -\left( \sqrt{1+\frac{m^2}{n^2}\sinh^2 \eta } -\cosh \eta
\right)^{\frac{3}{2\delta +3}}}{\left(
\sqrt{1+\frac{m^2}{n^2}\sinh^2 \eta} -\cosh \eta
\right)^{\frac{3}{2\delta +3}}}. \label{sol3}
\end{equation}
Knowing the value of the constants we can obtain the total energy
corresponding to the solution. It can be checked that
\begin{equation}
E_{m,n}=(2\pi)^2 4 \cdot 2^{1/4} \sqrt{|m||n|(|m|+|n|)}.
\label{energy4}
\end{equation}
Of course, more complicated dielectric function will bring us to
more complex solution. We would like to mention only two functions
which, in our opinion, lead to quite nice solutions. Namely, for
\begin{equation}
\sigma_1 (\vec{n})=\frac{a}{(1-n^3)^3} e^{-a\frac{3}{2} \left(
\frac{1+n^3}{1-n^3}\right)} \label{sigma1}
\end{equation}
we get
\begin{equation}
f^2=-\frac{1}{a} \ln \left[ \frac{1}{|m|-|n|}\left( |m| - |n|
\frac{\cosh \eta}{\sqrt{ \frac{n^2}{m^2} +\sinh^2 \eta }} \right)
\right], \label{sol_sigma1}
\end{equation}
whereas for
\begin{equation}
\sigma_2 (\vec{n})= \frac{8}{(1-n^3)^3} \left[ \frac{\sinh \left(
\frac{1+n^3}{1-n^3} \right)}{\cosh^2 \left( \frac{1+n^3}{1-n^3}
\right)} \right]^{\frac{3}{2}} \label{sigma2}
\end{equation}
one finds
\begin{equation}
f^2 ={\rm arcosh} \left[ \frac{|m|-|n|}{ \left( |m| - |n|
\frac{\cosh \eta}{\sqrt{ \frac{n^2}{m^2} +\sinh^2 \eta }} \right)
} \right]. \label{sol_sigma2}
\end{equation}
Both solutions correspond to the same constants $k_1, k_2$ and
possess identical total energy given by (\ref{energy4}).
\\
Let us now calculate the Hopf index for previously obtained field
configurations . It will be done for the solution (\ref{sol3}) but
the presented procedure, based on \cite{aratyn}, can be easily
repeated in case of solutions (\ref{sol_sigma1}),
(\ref{sol_sigma2}) as well. We introduce two additional functions
\begin{equation}
g_1^2= \left( \left(\frac{|m|}{|n|} -1\right)
\sqrt{\frac{n^2}{m^2} +\sinh^2 \eta} \right)^{\frac{3}{2\delta
+3}} -\left( \sqrt{1+\frac{m^2}{n^2}\sinh^2 \eta } -\cosh \eta
\right)^{\frac{3}{2\delta +3}}  \label{g1}
\end{equation}
and
\begin{equation}
g_2^2= \left( \sqrt{1+\frac{m^2}{n^2}\sinh^2 \eta} -\cosh \eta
\right)^{\frac{3}{2\delta +3}}. \label{g2}
\end{equation}
Then we can define functions $\Phi_i$, $i=1..4$
\begin{equation}
\Phi_{\left( ^1 _2 \right)}=\left(
\frac{g_1}{\sqrt{g_1^2+g_2^2}}\right) \times \left( ^{\cos m\xi}
_{\sin m \xi}  \right) \label{topPhi1}
\end{equation}
and
\begin{equation}
\Phi_{\left( ^3 _4 \right)}=\left(
\frac{g_2}{\sqrt{g_1^2+g_2^2}}\right) \times \left( ^{\cos n\phi}
_{-\sin n \phi}  \right), \label{topPhi3}
\end{equation}
which are connected with primary unit vector field by the relation
$n_i = Z^{\dagger} \sigma_i Z$, where
\begin{equation}
Z=\left(
\begin{array}{c}
Z_1 \\
Z_2
\end{array}
\right), \; \;
\; Z^{\dagger} =( Z_1^*, Z_2^*) \label{parametrHopf}
\end{equation}
and
\begin{equation}
Z_1=\Phi_1+i\Phi_2, \; \; \; Z_2=\Phi_3 +i\Phi_4.
\label{parametrZ}
\end{equation}
The crucial point is to find Abelian vector field which defines
the field tensor $H_{ij}=\partial_i A_j-\partial_j A_i$. It can be
achieved for the following potential
\begin{equation}
A_i = \frac{i}{2} (Z^{\dagger} \partial_i Z -\partial_i
Z^{\dagger} Z). \label{ab-potential}
\end{equation}
The Hopf index is defined in the standard way as
\begin{equation}
Q_H =\frac{1}{4\pi^2} \int d^3x \vec{A} \cdot \vec{B},
\label{indexHopf}
\end{equation}
where 'magnetic' field $\vec{B}=\nabla \times \vec{A}$. After
integration we find that
\begin{equation}
Q_H=\frac{nm}{2} \left[(\Phi^2_1+\Phi_2^2)^2 -
(\Phi_3^2+\Phi_4^2)^2 \right]_0^{\infty} =-mn. \label{hopf}
\end{equation}
For non-zero $m,n$ numbers the presented solutions possess the
non-trivial Hopf index. We would like notice that the famous
Vakulenko-Kapitansky non-equality \cite{vakulenko}, also in the
case where $O(3)$ symmetry is explicitly broken, is fulfilled. In
fact, we find that
\begin{equation}
E_{m,n} \geq (2\pi)^2 4\cdot 2^{\frac{3}{4}} |Q_H|^{\frac{3}{4}}.
\label{non-eq}
\end{equation}
One can see that we reproduce the result recently found in the
symmetric case \cite{aratyn}.
\section{\bf{ Conclusions}}
In the present paper $O(3)$ symmetry breaking models built of the
unit vector field leaving in the four dimensional Minkowski
space-time have been investigated. In general, the Lagrangian
consists of two terms. The first one, global $O(3)$ symmetric,
depends on the $H_{\mu \nu}^2$ is multiplied by a dielectric
function $\sigma (\vec{n})$. This function ensures the explicit
symmetry breaking on the Lagrangian level. \\
Let us summarized the achieved results.
\\
First of all, it has been proved that such models are integrable
in this sense that infinite number of the conserved currents can
be constructed. As we have mentioned it before, integrability of
models with broken global symmetry is rather exceptional effect if
we compare it with the standard theories with solitons. Usually,
integrability as well as appearance of solitons requires the
maximal possible symmetry.
\\
Secondly, the special case, where the $O(3)$ symmetric term is
chosen in the form to circumvent the famous Derick scaling
argument for non-existence of stable solitons, has been analyzed
in details. We have found a family toroidal solutions (in general
given by the integral), label by two integer numbers $m,n$. The
total energy has been also found. It is given by these numbers and
the first integration constant i.e. by the asymptotic value of the
field. For some particular dielectric functions the exact
solutions have been obtained. For the most interesting case where
the dielectric function has a form motivated from non-Abelian
color dielectric model \cite{my}, the generalized
Aratyn-Ferreira-Zimerman solutions have been presented. We have
also proved that these field configurations possess non-trivial
topology and can be classified by means of the Hopf index. Quite
intriguing, the energy-charge inequality takes identical form as
in the symmetric case.
\\
The main result of our work is that toroidal Hopf solitons can
appear in the model with broken global $O(3)$ symmetry. It should
be stressed that the $O(3)$ global symmetry is not essential for
existence of hopfions. \\
It is also worth to notice that analyzed here pattern of the
symmetry breaking (i.e. via the dielectric function) is a very
special one. The breaking of the symmetry is reflected merely in
the shape of the soliton. In spite of the fact that the form of
the soliton solutions strongly depends on the dielectric function
and can be really complicated, they all lead to the same total
energy and Hopf index as in the symmetric theory. Due to that our
model can be understood as the model where the $O(3)$ symmetry
breaking occurs in the minimal way.
\\
The existence of topological toroidal solutions in the model with
the explicit broken symmetry is very encouraging in the context of
the effective model for the low energy gluodynamics. As we have
mentioned, there are some numerical \cite{wipf} as well as
theoretical work \cite{niemi2}, \cite{sanchez1}, \cite{my} which
suggest that such effective model should break global $O(3)$
symmetry. In fact a model model with symmetry breaking dielectric
function has been recently proposed \cite{my}. Because of the fact
that symmetry breaking in this model occurs in the same manner as
considered here one can suppose that also for the model \cite{my}
the influences from the breaking terms may be insignificant. It
would be very interesting to investigate it more precisely. We
plan to address this problem in our proceeding paper.
\\ \\
This work is partially supported by Foundation for Polish Science
and ESF "COSLAB" programme.

\end{document}